# Large Spin Anomalous Hall Effect in $L1_0$-FePt: Symmetry and Magnetization Switching

Takeshi Seki[1,2,*], Satoshi Iihama[3], Tomohiro Taniguchi[4], and Koki Takanashi[1,2]

[1] *Institute for Materials Research, Tohoku University, Sendai 980-8577, Japan*

[2] *Center for Spintronics Research Network, Tohoku University, Sendai 980-8577, Japan*

[3] *WPI Advanced Institute for Materials Research, Tohoku University, Sendai 980-8577, Japan*

[4] *National Institute of Advanced Industrial Science and Technology (AIST), Spintronics Research Center, Tsukuba, 305-8568, Japan*

*Author to whom correspondence should be addressed.

Electronic mail: go-sai@imr.tohoku.ac.jp

**Abstract**

We quantitatively evaluate a spin anomalous Hall effect (SAHE), generating spin angular momentum flow (spin current, $J_s$), in an $L1_0$-FePt ferromagnet by exploiting giant magnetoresistance devices with $L1_0$-FePt / Cu / Ni$_{81}$Fe$_{19}$. From the ferromagnetic resonance linewidth modulated by the charge current ($J_c$) injection, the spin anomalous Hall angle ($\alpha_{\text{SAH}}$) is obtained to be $0.25 \pm 0.03$. The evaluation of $\alpha_{\text{SAH}}$ at different configurations between $J_c$ and magnetization enables us to discuss the symmetry of SAHE and gives the unambiguous evidence that SAHE is the source of $J_s$. Thanks to the large $\alpha_{\text{SAH}}$, we demonstrate the SAHE-induced magnetization switching.

(99 words)

High efficient conversion from $J_c$ to $J_s$ and vice versa is the key for spintronics to enhance device performance and to provide with multi-functionality. The most promising way to convert from $J_c$ to $J_s$ is to exploit spin Hall effect (SHE) [Refs. 1, 2]. SHE is a relativistic effect expressed as

$$\mathbf{J}_s = (\hbar/2e)\alpha_{SH}[\hat{\sigma} \times \mathbf{J}_c], \qquad (1)$$

where $\alpha_{SH}$ is the spin Hall angle, $e$ (< 0) is the electric charge of an electron and $\hat{\sigma}$ is the quantization axis of electron spin. The experimental studies on SHE have first been done in nonmagnetic semiconductors [Refs. 3,4] followed by nonmagnetic metals with large spin orbit coupling parameters [Ref. 5]. As in SHE of nonmagnets (NM), one can expect the charge-spin conversion in a ferromagnet (FM). In the first stage, the conversion from $J_s$ to $J_c$ in FM was reported [Refs. 6-8], followed by the theory of spin anomalous Hall effect (SAHE) [Ref. 9] and experiments generating $J_s$ [Refs. 10-13]. The most apparent difference between SHE in NM and SAHE in FM is that the charge current appears in the transverse direction due to the anomalous Hall effect (AHE) [Ref. 14], $J_c^{AHE}$, which is defined as

$$\mathbf{J}_c^{AHE} = \alpha_{AH}[\hat{p} \times \mathbf{J}_c], \qquad (2)$$

where $\alpha_{AH}$ is the anomalous Hall angle and $\hat{p}$ is the quantization axis of local spin in FM. This transverse $J_c^{AHE}$ is spin-polarized by the factor of $\zeta$ that refers to the polarization of AHE [Ref. 9]. Although this $J_c^{AHE}$ gives rise to $-J_c^{AHE}$ in the open circuit condition, $-J_c^{AHE}$ is spin-polarized by the factor of $\beta$ representing the spin polarization factor in the longitudinal $J_c$ flow. Consequently, $J_s$ originating from SAHE can be written as [Ref. 9]

$$\mathbf{J}_s^{SAHE} = (\hbar/2e)(\zeta - \beta)\alpha_{AH}[\hat{p} \times \mathbf{J}_c], \qquad (3)$$

where $(\zeta - \beta)\alpha_{AH}$ corresponds to spin anomalous Hall angle ($\alpha_{SAH}$). Equation (3) means that $J_s^{SAHE}$ depends on $\hat{p}$, *i.e.* **M**. If $\hat{p}$ // $\mathbf{J}_c$, $J_s^{SAHE}$ becomes zero. In contrast to the reports on SAHE [Refs.

10-14], the recent studies found the various phenomena different from SAHE such as SHE [Ref. 15] and an interface scattering effect [Refs. 16,17]. Therefore, the identification of SAHE requires the elimination of these other effects. Another important issue confronted with SAHE is its low conversion efficiency. Since SAHE has benefits that SHE does not have, *e.g.* the ability of reorientation of antidamping torque along $\hat{p}$ and resultant zero-field magnetization switching [Ref. 9], a FM exhibiting large $\alpha_{SAH}$ is first essential for practical applications.

In this study, we focus on a ferromagnetic metal $L1_0$-FePt because large AHE has been reported for $L1_0$-FePt [Ref. 18]. In order to evaluate $\alpha_{SAH}$ quantitatively, we carried out the ferromagnetic resonance (FMR) measurement for the giant magnetoresistance (GMR) device composed of $L1_0$-FePt ($d^{FePt}$ = 30 nm) | Cu (3 nm) | $Ni_{81}Fe_{19}$ (Permalloy: Py, $d^{Py}$ =2 nm), in which the $L1_0$-FePt has the in-plane uniaxial magnetic anisotropy. **Fig. 1** depicts the concept of the present experiment and the measurement configurations. The application of dc current (corresponding to $J_c$) flowing in the $L1_0$-FePt layer is converted into $J_s^{SAHE}$, and $J_s^{SAHE}$ interacts with $M$ of Py (its unit vector is defined as **m**), resulting in the modification of FMR linewidth due to the enhanced or reduced magnetization damping of Py. A key of this study is that the strong in-plane uniaxial magnetic anisotropy of $L1_0$-FePt enables us to utilize two different relative configurations between $J_c$ and $M$ of FePt (its unit vector is defined as **p**). As shown in **Fig. 1(a)**, the orthogonal configuration of $\mathbf{J}_c \perp \mathbf{p}$ corresponds to the evaluation of SAHE. Other possible effects such as SHE and interface effect can be identified in the parallel configuration of $\mathbf{J}_c // \mathbf{p}$. We fabricated the rectangular-shaped devices with the orthogonal and parallel configurations on the identical substrate (**Figs. 1(b) and 1(c)**). Together with dc current ($I_{dc}$), the radiofrequency current ($I_{rf}$) was applied to generate an oscillating magnetic field, which excited FMR in the Py. The length of GMR device was fixed at 10 μm while the width (*w*) of

Page 4

device was varied: $w$ = 1 μm or 4 μm.

The thin films were grown on an SrTiO$_3$ (110) single crystal substrate with the stack of SrTiO$_3$ subs. || FePt (30) | Cu (2) | Py (3) | Al-O (10) (in nanometer). All the layers except FePt were grown at ambient temperature. The FePt (110) layer was epitaxially grown on the SrTiO$_3$ substrate using an ultrahigh vacuum magnetron sputtering system at 450 ºC. After the FePt deposition, the sample was transferred to the ion beam sputtering (IBS) chamber. After cleaning the FePt surface by soft Ar ion milling, the Cu | Py | Al-O layers were deposited on the FePt employing the IBS. The film structure was characterized using the x-ray diffraction (XRD) with Cu-$K\alpha$ radiation. The alloy compositions of FePt and Py were determined to be Fe$_{54}$Pt$_{46}$ and Ni$_{81}$Fe$_{19}$, respectively, by electron probe x-ray microanalysis. Magnetization was measured at room temperature using the vibrating sample magnetometer. The thin films were patterned into the rectangular shapes through the use of electron beam lithography and Ar ion milling. Finally, the Au | Cr contact pads with the co-planar waveguide (CPW) shape were formed. For the FMR measurement, the rf power of 5 dBm with a fixed $f_0$ was applied to the signal line of CPW from a signal generator, which induced the transverse rf magnetic field. When the FMR condition of Py was satisfied with sweeping $H_{ext}$, the device resistance ($R(t)$) oscillated through the GMR effect. As a result, the applied $I_{rf}(t)$ ($= I \cos(2\pi f_0 t)$) and the oscillating $R(t)$ ($\propto \cos(2\pi f_0 t)$) generated the rectification dc voltage ($V_{dc}$), which was detected by the lock-in amplifier. For the FMR measurement with the $I_{dc}$ application, the sourcemeter was connected to the dc port of bias-T.

Let us first explain the structure and the magnetic properties of GMR tri-layer. The out-of-plane x-ray diffraction profile (**Fig. 2(a)**) shows only the 220 diffractions of FePt, Cu and Py while in the in-plane XRD (**Fig. 2(b)**) only the 00$l$ diffractions of FePt, Cu and Py are observed. The in-plane $\phi$ scan for FePt 200 diffraction (**Fig. 2(c)**) exhibits the two-fold rotational symmetry.

Therefore, the GMR tri-layer was grown with the epitaxial relationship of $(110)_{SrTiO_3}$ || $(110)_{FePt}$ || $(110)_{Cu}$ || $(110)_{Py}$, $[001]_{SrTiO_3}$ || $[001]_{FePt}$ || $[001]_{Cu}$ || $[001]_{Py}$. In addition, the appearance of FePt 001 and 003 superlattice peaks in **Fig. 2(b)** indicates that $L1_0$-ordered FePt was formed. The crystal orientation relationship between $SrTiO_3$ and $L1_0$-FePt is schematically shown in **Fig. 2(d)**. **Fig. 2(e)** represents the magnetization curves with $H_{ext}$ applied along the in-plane [001] and $[1\bar{1}0]$ directions and the out-of-plane [110] direction. It is obvious that the $L1_0$-FePt layer has the strong in-plane uniaxial magnetic anisotropy along the in-plane [001] direction. As in the case of $L1_0$-FePt layer, the Py also exhibits the non-negligible in-plane magnetic anisotropy along the [001]. The two-step behavior of magnetization switching is observed in the in-plane [001] magnetization curve, indicating the induction of coercivity difference between $L1_0$-FePt and Py.

**Fig. 3(a)** shows the current-in-plane (CIP) GMR curves for the orthogonal configuration-device with $w$ = 4 μm at the magnetic field angles ($\theta_H$), defined in **Fig. 1(b)**, of 0º, 15º, and 90º. The detailed $\theta_H$ dependence of MR curve for the orthogonal and parallel configuration-devices are given in **Supplemental Materials** [Ref. 19]. The orthogonal configuration was achieved for $\theta_H$ = 0º by increasing $H_{ext}$. On the other hand, the full GMR curve was obtained when $H_{ext}$ was swept at $\theta_H$ = 90º. These MR curves indicate that **m** is easily aligned along $\mathbf{H}_{ext}$ while **p** acts as a fixed polarizer, resulting in the orthogonal configuration at small $\theta_H$. **Figs. 3(b) and 3(c)** show the FMR spectra for the orthogonal configuration-device at $I_{dc}$ = 4 mA and -4 mA, respectively. $\theta_H$ was set at -10º and $f_0$ was 4 GHz. $V_{dc}$ was generated by the rectification effect through the CIP-GMR effect, and the resonance peak comes from the FMR of Py. The experimental spectra were fitted by the Lorentzian and anti-Lorentzian functions together with the linear background for taking into account the slight change in the device resistance with $H_{ext}$ (see [Ref. 19] for details of fitting). One can see the clear variation in the spectral linewidth ($\Delta H_{Res}$) between $I_{dc}$ = 4 mA and -4 mA. The $\Delta H_{Res}$ versus $I_{dc}$ for $\theta_H$ = -10° and +15° is plotted in **Figs. 3(d) and 3(e)**, respectively. The linear variations are observed, and it slope depends on $\theta_H$. It is noted that the odd-functional behavior of $\Delta H_{Res}$ against $I_{dc}$ is not explained by the spin transfer torque (STT) of GMR stack because the sign of STT, *i.e.* damping or anti-damping, in the CIP-GMR device is not determined by the $I_{dc}$ direction. **Fig. 3(f)** summarizes the modulation linewidth, $d\Delta H_{Res} / dI_{dc}$, as a function of $\theta_H$ for the orthogonal configuration-device. The experimental data were well fitted by the following equation:

$$\frac{d\Delta H_{Res}}{dI_{dc}} = \alpha_{SAH} \frac{\hbar}{e} \left(\frac{C^{FePt}}{wd^{FePt}}\right) \cdot \left(\frac{2\pi f_0}{\gamma \cos(\varphi^{Py} - \theta_H)}\right) \cdot \left(\frac{\sin \varphi^{Py}}{\mu_0 M_s^{Py} d^{Py} (H_{XX} + H_{YY})}\right), \quad (4)$$

where $C^{FePt}$ is the shunt ratio of the current flowing in the $L1_0$-FePt layer to the total $I_{dc}$, $\gamma$ is the gyromagnetic ratio, $\varphi^{Py}$ is the angle of **m** respective to $\mathbf{J}_c$, $M_s^{Py}$ is the saturation magnetization of Py, $H_{XX(YY)}$ represents the in-plane (out-of-plane) effective field for Py (see [Ref. 19] for the derivation of Eq. (4)). We assumed that **p** is pinned along the direction orthogonal to $\mathbf{J}_c$. The values of $\gamma$, $\varphi^{Py}$, $M_s^{Py}$ and $H_{XX(YY)}$ were determined by the $\theta_H$ dependence of FMR spectra without $I_{dc}$ application, and $C^{FePt}$ = 0.315 was calculated considering the parallel circuit (see [Ref. 19]). As a result, $\alpha_{SAH}$ for the orthogonal configuration-device was obtained to be 0.25 ± 0.03. Let us discuss the magnitude of $\alpha_{SAH}$ here. This $\alpha_{SAH}$ is significantly larger than - 0.14 ± 0.05 for CoFeB [Ref. 12]. As described above, $\alpha_{SAH}$ is given by $(\zeta - \beta)\alpha_{AH}$. $\alpha_{AH}$ of FePt was evaluated to be 0.03 by measuring AHE (see [Ref. 19]). The previous transport experiments using $L1_0$-FePt suggest $\beta \sim 0.4$ at most [Ref. 20]. This means that large $\zeta$ = 8.7 is required in order to explain the large $\alpha_{SAH}$. We note that the parameter $\zeta$ is defined as a ratio of the spin current to the transverse charge current by AHE. Therefore, if the spin-up and spin-down electrons are scattered to the opposite direction, $\zeta$ can be larger than one. The parameter $\zeta$ was introduced in Ref. 9 phenomenologically, and an evaluation of its value by, for example, first-principles calculation has not been reported yet. Our result, however, suggests that even a FM with a small AHE ($\alpha_{AH}$) has a possibility to be a high-efficient $J_s$ source by SAHE.

As mentioned above, several origins of $J_s$, other than SAHE, have been proposed in CIP-GMR structures. For example, the inverse SHE of Co was reported in Co | Cu | $Y_3Fe_5O_{12}$ [Ref. 14], where the spin polarization is geometrically determined and is independent from **M**. We confirm that such *M*-independent SHE does not exist in our sample by measuring the FMR in the parallel configuration. Note that SAHE becomes zero in the parallel configuration whereas the *M*-independent SHE is finite if exists. However, no clear $\theta_H$ dependence of $d\Delta H_{Res} / dI_{dc}$ was seen for the parallel configuration-device (**Fig. 3(g)**, the FMR spectra are given in [Ref. 19]), indicating that the *M*-independent SHE is negligible in our experiment. Another possible source of $J_s$ is the interface scattering effect [Ref. 16]. Recently, the interface-induced spin orbit torque has also been observed in Py (or CoFeB) | Ti | CoFeB [Ref. 17]. $J_s$ generated by the interface scattering effect provides two kinds of spin torques. One has the spin polarization identical to that of the *M*-independent SHE, and therefore, is negligible due to the same reason explained above. The other has the polarization along the direction perpendicular to the film plane, as in the case of the previous work [Ref. 17]. We note that the average of this torque over the precession of **m** becomes zero. Therefore, this torque does not affect $\Delta H_{Res}$. In summary, several phenomena of $J_s$ generation reported in the previous works are negligibly small or do not affect our measurement, and we can conclude that the modulation of $\Delta H_{Res}$ is solely due to the SAHE.

Let us move to the magnetization switching experiment. We characterized the static magneto-transport properties under the $I_{dc}$ application for the orthogonal configuration-device with $w$ = 1 μm. The minor GMR curves at $I_{dc}$ = 0 mA and ± 4.8 mA are shown in **Figs. 4(a) and 4(b)**, respectively, where $\theta_H$ was set at 60º in order to increase the projection component of **m** to **p**. This projection component is inevitable for the deterministic switching. On the other hand, this $\theta_H$ led to the partial (not full) magnetization switching of Py. The remarkable loop shift was observed by applying $I_{dc}$, which is attributable to the SAH torque. The switching experiment was carried out using the measurement sequence schematically shown in **Fig. 4(c)**. **Fig. 4(d)** demonstrates that the $I_{dc}$ application can switch the magnetization direction of Py. This switching behavior is similar to conventional spin orbit torque switching using nonmagnetic materials [Refs. 21-23] and topological insulators [Refs. 24,25]. The current density ($J$) for switching is shifted with $H_{ext}$. We achieved the small switching current of $|J| < 10^7$ A/cm$^2$ at $H_{ext}$ = 325 Oe. This value is consistent with the small critical switching current density of ~ $3.1 \times 10^6$ A/cm$^2$ estimated theoretically with the experimental $\alpha_{SAH}$ (see [Ref. 19]). We also examined the current-induced switching at the condition where **p** has the opposite polarity and the $\theta_H$ dependence of current-induced switching (see [Ref. 19]), which were well interpreted within the framework of SAHE. One may think the contribution of current-induced Oersted field ($H_{Oe}$) for switching. We calculated $H_{Oe}$ at the switching current and $H_{Oe}$ was estimated to be ~10 Oe at 1.8 mA, which is smaller than the coercivity of Py (~25 Oe) (see [Ref. 19]). Therefore, we consider that the SAH torque plays the major role for magnetization switching. In contrast to the orthogonal configuration, the switching behavior was not observed for the parallel configuration-device (see [Ref. 19]), again indicating negligible SHE and interface scattering effect. We conclude this is the first demonstration of SAHE-induced magnetization switching. In addition to the evaluation of linewidth modulation and the demonstration of switching, we examined the FMR lineshape analysis and the control experiment using the Cu | Py | Al-O device was done, which are given in [Ref. 19]. Those additional experiments support our conclusion of large SAHE in $L1_0$-FePt.

A great advantage for SAHE is the possibility of zero-field switching of perpendicular **m** by employing a SAHE polarizer with **p** tilted normal to the device plane. Since $L1_0$-FePt (111) with tilted magnetic anisotropy is exploited as a tilted polarizer [Ref. 26], we believe that $L1_0$-FePt showing large

Page 9

SAHE becomes the key material to tackle the essential difficulty the spin orbit torque devices are confronted with.

**Acknowledgements**

The authors thank K. Uchida, T. Kikkawa, and S. Takahashi for valuable discussions. I. Narita provided technical support during the structural characterization. This work was supported by the Grant-in-Aid for Scientific Research B (16H04487), Grant-in-Aid for Challenging Research (Exploratory, 18K19012), Grant-in-Aid for Scientific Research (S) (JP18H05246) and Innovative Area "Nano Spin Conversion Science" (26103006) as well as the Research Grant from the TEPCO Memorial Foundation. The device fabrication was partly carried out at the Cooperative Research and Development Center for Advanced Materials, IMR, Tohoku University.

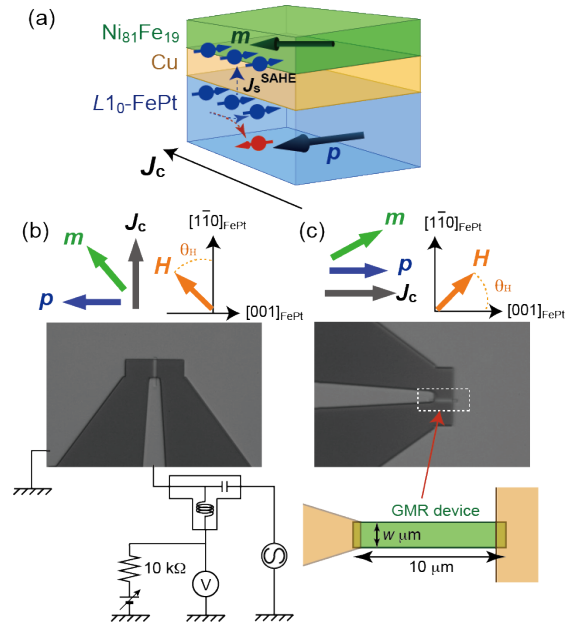

**Figure 1 (a)** Schematic illustration of SAHE in $L1_0$-FePt (30 nm) | Cu (3 nm) | Py (2 nm), where **p** and **m** are the unit vectors of $L1_0$-FePt and Py magnetizations, respectively. $\mathbf{J_c}$ denotes the charge current flow whereas $\mathbf{J_s}^{SAHE}$ denotes the spin current originating from SAHE. **(b)** Optical microscope images of devices for the orthogonal configuration and **(c)** the parallel configuration together with the relationship between **p**, **m**, $\mathbf{J_c}$, and external magnetic field ($\mathbf{H_{ext}}$). $\theta_H$ was defined to be the angle from $\mathbf{J_c}$. The setup for FMR measurement and the enlarged illustration of GMR device are also depicted.

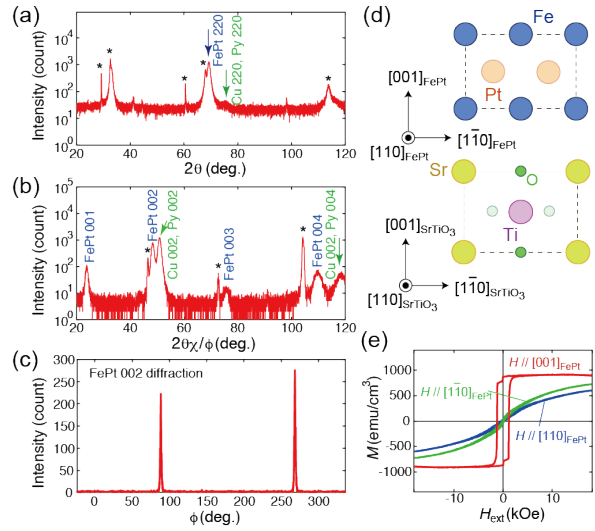

**Figure 2 (a)** Out-of-plane and **(b)** in-plane x-ray diffraction profiles for the $L1_0$-FePt (30 nm) | Cu (3 nm) | Py (2 nm) stack. The asterisks denote the reflections from the SrTiO$_3$ (110) substrate. **(c)** In-plane $\phi$ scan for FePt 200 diffraction. **(d)** Illustration of the epitaxial relationship between SrTiO$_3$ (110) and $L1_0$-FePt (110) planes. **(e)** Magnetization curves with $H_{ext}$ applied along the in-plane [001] (red line) direction, the in-plane $[1\bar{1}0]$ direction (green line), and the out-of-plane [110] direction (blue line).

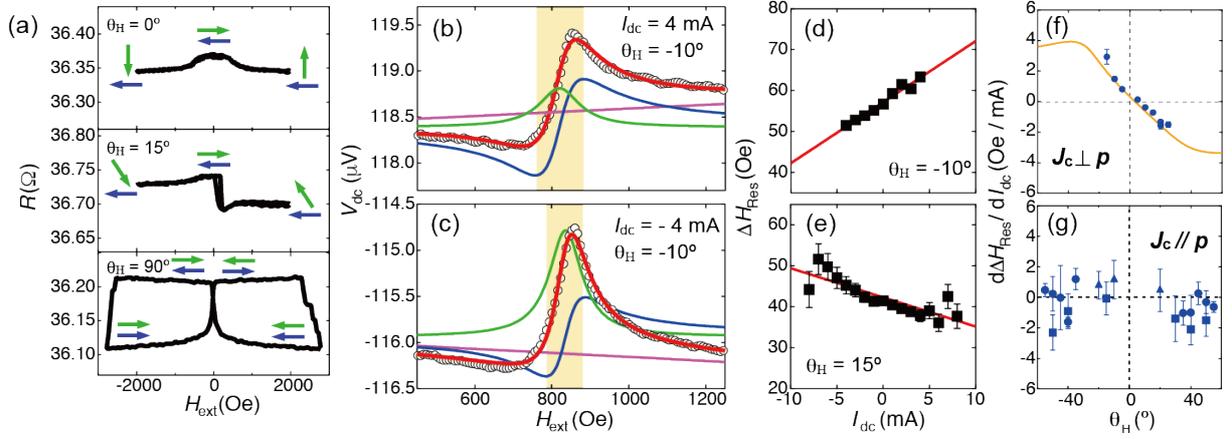

**Figure 3 (a)** Magnetoresistance curves for orthogonal configuration-device with $w = 4$ μm at $\theta_H = 0°$, 15°, and 90°. The blue and green arrows denote the $L1_0$-FePt and Py magnetizations, respectively. **(b)** FMR spectra for orthogonal configuration-device with $w = 4$ μm under the application of $I_{dc} = 4$ mA and **(c)** -4 mA. $\theta_H$ was set at -10° and $f_0$ was 4 GHz. The experimental data (black open circles) were fitted by the function (red curves) composed of lorentzian (green curves) and anti-lorentzian (blue curves) together with the linear background (purple lines). The orange-hatched areas are guides for eyes to see the linewidth modulation. **(d)** $\Delta H_{Res}$ versus $I_{dc}$ for $\theta_H = -10°$ and **(e)** $+15°$ for orthogonal configuration-device. **(f)** $d\Delta H_{Res} / dI_{dc}$ as a function of $\theta_H$ for the orthogonal configuration- and **(g)** parallel configuration-devices. The solid curve in **(f)** represents the fitting result using Eq. 4. In **(g)**, different three devices (denoted by the solid circles, triangles and squares) were evaluated.

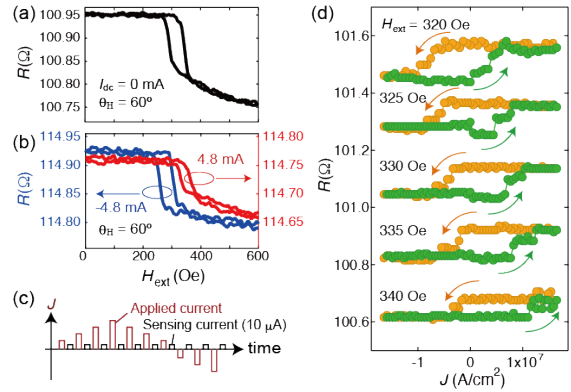

**Figure 4 (a)** Minor curves of CIP-GMR for orthogonal configuration-device with $w = 1$ μm at $\theta_H = 60°$ without applying $I_{dc}$ and **(b)** with $I_{dc} = 4.8$ mA (red curve) and $-4.8$ mA (blue curve). **(c)** Measurement sequence for the switching experiment. First, "Applied current" was injected into the device, then $R$ was measured by small "Sensing current" of 10 μA. **(d)** $R$ as a function of applied current density ($J$) for orthogonal configuration-device measured at various $H_{ext}$. For clarity, the loops are shifted vertically. The arrows denote the current sweep directions.